\PassOptionsToPackage{dvipsnames}{xcolor}

\documentclass[twocolumn, linenumbers, usenatbib]{mnras}
\usepackage{acronym}
\usepackage{xspace}
\usepackage[dvipsnames]{xcolor}
\usepackage[normalem]{ulem}
\usepackage{amsmath}
\usepackage{ragged2e}
\usepackage{booktabs}
\usepackage{multirow}
\usepackage{paralist}
\usepackage{hyperref}
\usepackage{amsmath,amssymb}
\usepackage{orcidlink}
\PassOptionsToPackage{breaklinks, plainpages=false, anchorcolor=cyan, linkcolor=magenta, citecolor=cyan, urlcolor=violet, bookmarks=false}{hyperref}

\newacro{imbh}[IMBH]{intermediate-mass black hole}
\newacro{bhns}[BHNS]{black hole neutron star}
\newacro{bbh}[BBH]{binary black hole}
\newacro{bh}[BH]{black hole}
\newacro{bns}[BNS]{binary neutron star}
\acrodef{FAR}[FAR]{false alarm rate}
\newacro{bf}[BF]{Bayes' factor}
\newacro{cbc}[CBC]{compact binary coalescence}
\newacro{ce}[CE]{Cosmic Explorer}
\acrodef{SNe}[SNe]{Supernova}
\newacro{da}[DA]{data analysis}
\newacro{et}[ET]{Einstein Telescope}
\newacro{eob}[EOB]{Effective-One-Body}
\newacro{fd}[FD]{frequency domain}
\newacro{gw}[GW]{gravitational-wave}
\newacro{gr}[GR]{general relativity}
\newacro{hm}[HM]{Higher mode}
\newacro{ifo}[IFO]{interferometer}
\newacro{imr}[IMR]{inspiral-merger-ringdown}
\newacro{im}[IM]{inspiral-to-merger}
\newacro{kagra}[KAGRA]{Kamioka Gravitational Wave Detector}
\newacro{ligo}[LIGO]{Laser Interferometer Gravitational-Wave Observatory}
\newacro{lso}[LSO]{Last Stable Orbit}
\newacro{lvc}[LVC]{LIGO-Virgo Collaboration}
\newacro{lvk}[LVK]{LIGO-Virgo-Kagra Collaboration}
\newacro{ns}[NS]{neutron star}
\newacro{nr}[NR]{numerical relativity}
\newacro{pn}[PN]{post-Newtonian}
\newacro{pe}[PE]{parameter estimation}
\newacro{psd}[PSD]{power spectral density}
\newacro{asd}[ASD]{amplitude spectral density}
\newacro{xg}[XG]{next-generation}
\newacro{jsd}[JSD]{jensen shannon divergence}
\newacro{sgrb}[sGRB]{short gamma ray burst}
\newacro{igwn}[IGWN]{international gravitational wave network}
\newacro{snr}[SNR]{signal-to-noise ratio}
\newacro{eos}[EoS]{equation of state}
\newacro{em}[EM]{electromagnetic}
\newacro{qnm}[QNM]{quasi-normal mode}

\newcommand{\nrsur}{\texttt{NRSur7dq4}\xspace}

\newcommand{\bilby}{\texttt{Bilby}\xspace}
\newcommand{\dynesty}{\texttt{dynesty}\xspace}

\title[Eccentricity in Disguise]{Eccentricity in Disguise? Insights from GW231123 and\\ Numerically Simulated Binary Black Hole Merger Signals}

\author[Koustav Chandra et al]{Koustav Chandra$^{1,2}$ \thanks{koustav.chandra@aei.mpg.de} Johann Fernandes$^{3}$, 
Akshita Mittal$^{4,5}$
Gregorio Carullo$^{6}$
\\
$^{1}$ Max Planck Institute for Gravitational Physics (Albert Einstein Institute), Am M\"uhlenberg 1, 14476 Potsdam, Germany\\
$^{2}$ Institute for Gravitation \& the Cosmos, Department of Physics, Penn State University, University Park, PA 16802, USA \\
$^{3}$ Indian Institute of Technology, Bombay, Mumbai-400076, India \\ 
$^{4}$ Gran Sasso Science Institute (GSSI), I-67100 L’Aquila, Italy \\
$^{5}$ INFN, Laboratori Nazionali del Gran Sasso, I-67100 Assergi, Italy \\
$^{6}$ School of Physics and Astronomy and Institute for Gravitational Wave Astronomy, University of Birmingham, Edgbaston, \\ Birmingham, B15 2TT, United Kingdom
}

\begin{document}
\maketitle





\date{\today}

\begin{abstract}

GW231123 is a gravitational-wave signal originating from the merger of a black hole binary with total mass $\sim 250 M_{\odot}$, the largest ever detected by the LIGO-Virgo-Kagra Collaboration.
Remarkably, under standard priors, the system features among the fastest-spinning binary components confidently measured in binary mergers, $ \chi_{1,2} \gtrsim 0.7$ at \(90\%\) one-dimensional credibility, according to the most accurate model employed.
As typical binary mergers result in remnants with $\chi \sim 0.7$, such spin values are challenging to obtain even from previous (hierarchical) mergers.
These inferred properties rely on waveform models lacking eccentric corrections in the merger-ringdown stage.
Here, we show that binaries retaining significant eccentricity up to merger can be misinterpreted as \textit{near-extremally} spinning when non-circular corrections are neglected.
Binary-agnostic ringdown analysis instead provides unbiased estimates of the remnant properties, provided that a robust estimate of the signal peak can be obtained.
We re-analyse GW231123 using available eccentric numerical-relativity catalogues, finding that although eccentric templates can provide a good fit to the data, quasi-spherical templates are still favoured.
Ringdown analyses confirm a secondary likelihood peak correlated with large eccentricity values, but improved eccentric models will be required to assess the reliability of this interpretation.
Finally, analysing GW231123 under population-informed parametric priors confirms the exceptional nature of this event within the current black hole binary population.

\end{abstract}

\section{Introduction}\label{sec:introduction}

Statistically significant, short-duration (\(\lesssim 0.02\,\mathrm{s}\)) \ac{gw} events such as GW190521~\citep{LIGOScientific:2020iuh} and GW231123\_135430 (henceforth GW231123)~\citep{LIGOScientific:2025rsn} are consistent with mergers of two \acp{bh}, resulting in a remnant \ac{bh} with mass $ M_f > 100\,M_{\odot}$.
Their properties place the progenitor binary components in the tail of the mass distribution of the \ac{bbh} population observed by the Advanced LIGO--Virgo--Kagra network~\citep{LIGOScientific:2014pky, VIRGO:2014yos, KAGRA:2013rdx, LIGOScientific:2025pvj}, and their remnants confidently within the (observationally elusive) \ac{imbh} mass range~\citep{Greene_2020}.

The large total mass of these binaries indicates that they likely formed in a dense environment, where dynamical interactions are frequent~\citep{10.3389/fspas.2020.00038, Gerosa:2021mno}. 
Such environments can impart several distinctive properties to the merging binary: higher masses due to hierarchical mergers, significant asymmetries in component masses, intrinsic rotation rates (spins) misaligned with respect to the orbital plane (giving rise to spin-precession), and, most importantly for our work, non-negligible eccentricity \citep{10.1093/mnras/stad3048, 2014ApJ...781...45A, Samsing_2018, Rodriguez_2018, Zevin_2019, DallAmico:2023neb}. 
By contrast, binaries formed in isolation are expected to be more mass-symmetric, with low-aligned spins and negligible eccentricity by the time they enter the ground-based detector band, owing to efficient loss of eccentricity during inspiral via \ac{gw} emission~\citep{Peters:1963ux}.
Therefore, detection of non-negligible eccentricity can provide a clear and robust signature of the dynamical origin of these signals.

Owing to the limited low-frequency sensitivity of current interferometers, the vast majority of the measurable power in these massive signals is concentrated in the few cycles encompassing the merger-ringdown coalescence stage.
The efficient loss of eccentricity during a \ac{gw}-driven inspiral implies that retaining significant binary eccentricity up to the merger stage requires rare interactions with the surrounding environment or formation by close encounter.
For this reason, standard analyses assume quasi-spherical \ac{bbh} orbits, allowing for generic spin orientations but negligible eccentricity~\citep{LIGOScientific:2020iuh, Nitz:2021zwj, Estelles:2021jnz, Olsen:2021qin}.
Nevertheless, more general orbits for GW190521 have also been explored (e.g.~\citep{CalderonBustillo:2020xms}), including a non-precessing binary on an eccentric orbit circularising before merger~\citep{Romero-Shaw:2020thy, Iglesias:2022xfc, Gupte:2024jfe}, which does not yield significant evidence for eccentricity.
Instead, an analysis based on a sparse grid of \ac{nr} simulated \ac{bbh} signals has shown that GW190521 is consistent with the merger signal of an eccentric, spin-precessing \ac{bbh}~\citep{Gayathri:2020coq}. 
Finally, a dynamical capture~\citep{Gamba:2021gap} hypothesis was shown to be preferred by the data, while assuming a semi-analytical waveform model employing non-spinning components and a quasi-circular merger ringdown~\footnote{Throughout this work, we use the term \emph{quasi-circular} to denote non-eccentric, non-precessing binaries, while \emph{quasi-spherical} refers to non-eccentric binaries with spin-induced precession.}. 
This result has recently been validated when including spin-aligned binary components~\citep{Lange:2026eqx, Pompili:2026yxq}, albeit with weaker evidence.

For GW231123, an analysis with the eccentric and spin-precessing model \texttt{TEOBResumS-Dalí}~\citep{Chiaramello:2020ehz, Nagar:2020xsk, Nagar:2024dzj, Nagar:2024oyk, Gamba:2024cvy}, under a bounded orbital configuration, showed evidence in support of eccentricity~\citep{Jan:2025zcm, Malagon:2026uev}.
It also showed that inference with an eccentric, non-precessing waveform could yield a confident non-zero eccentricity measurement when spin precession is neglected. 
On the contrary, it has also been shown that for these types of signals, the inferred precession could actually stem from misinterpretation of the signal's high eccentricity~\citep{CalderonBustillo:2020xms, Divyajyoti:2025cwq, Romero-Shaw:2022fbf}.
For this signal, the dynamical capture hypothesis is disfavoured compared to a quasi-spherical one~\citep{Lange:2026eqx, Pompili:2026yxq}.
All these analyses relied on a quasi-circular merger-ringdown model, which might yield significant systematic uncertainties for such short merger-dominated signals.
The complete resolution of such a variety of statements will require a robust statistical analysis based on a waveform model that incorporates spin precession along generic non-circular orbits (both bounded-eccentric and dynamically-bounded), including non-circular corrections in the merger-ringdown portion.
This will be of paramount importance for disentangling different formation channels contributing to the mass spectrum around the pair-instability mass gap~\citep{Baumgarte2025,Li2025,Ray2025,Liu2025,Angeloni2026}.

The short duration of these signals and the consequent challenge in interpreting their properties have also given rise to other exotic interpretations.
For instance, GW190521 has been argued to be consistent with head-on collisions of Proca stars~\citep{CalderonBustillo:2020fyi, CalderonBustillo:2022cja} or the collapse of cosmic strings~\citep{Aurrekoetxea:2023vtp, Cuceu:2025fzi} or a high-mass disk rotating a \ac{bh}~\citep{Shibata:2021sau}. 
These exotic scenarios are far less astrophysically plausible than a \ac{bbh} merger hypothesis, while also not reaching comparable likelihoods of astrophysical \ac{bbh} mergers~\citep{Gayathri:2020coq, LIGOScientific:2025rsn}.
For these reasons, we will not consider them here.
We also note that the short duration makes these signals potentially sensitive to sub-threshold non-Gaussian features that deviate from the underlying assumption of a Gaussian stationary background~\citep{Ray:2025rtt}, although no evidence of biases induced by such effects on the actual events has yet been reported~\citep{Bini:2026kwz}. Additionally, waveform reconstructions by \citet{Chatterjee:2025avc} found that machine-learning-based reconstructions of GW231123 achieve higher overlaps with unmodelled reconstructions than with posterior-supported waveforms obtained under quasi-spherical hypothesis, potentially indicating unmodelled source physics, such as eccentricity, or residual waveform systematics.

In this paper, we investigate the impact of eccentricity on another peculiar but informative, and hard-to-explain property~\citep{Passenger:2025acb,Popa2025,Delfavero2025,Stegmann2025,Croon2025} of GW231123: its very large progenitors' spin magnitude.
We analyse GW231123 using state-of-the-art \ac{nr} waveforms with initial eccentricity \(e_0 \gtrsim 0.5\), focusing on configurations where the post-peak signal is substantially affected.

We proceed in three steps. 
\textit{First}, we quantify the extent to which eccentric signals can be misidentified as non-eccentric by comparing against available \ac{nr} waveforms using the quasi-spherical \nrsur{} model~\citep{Varma:2019csw}. 
\textit{Second}, we assess the impact of such mismodelling on parameter inference through both full Bayesian analyses and targeted post-peak (ringdown) studies of a subset of numerical simulations. 
\textit{Finally}, we analyse the GW231123 data using a suite of eccentric \ac{nr} waveforms to test whether such configurations provide a viable description of the observed signal.

We find that short-duration signals from eccentric binaries with modest component spins aligned with the orbital angular momentum ($\chi_{1,2} \lesssim 0.5$) can be misinterpreted as originating from quasi-spherical systems with large in-plane spins ($\chi_1 \gtrsim 0.8$ at $90\%$ credibility).
Waveform mismatches are partially absorbed through shifts in intrinsic parameters, resulting in inferred spin orientations that are strongly misaligned with the orbital angular momentum and opposite to the true configuration. 
However, for \acp{snr} $\gtrsim 20$, model-selection diagnostics can confidently distinguish between eccentric and quasi-spherical hypotheses, indicating that this apparent degeneracy arises at the level of parameter estimation rather than from an intrinsic indistinguishability of the underlying signals. 
In contrast, post-peak analyses at sufficiently late times recover remnant parameters consistent with Kerr expectations, as the late-time ringdown is mainly determined by the remnant's mass and spin.
When applied to GW231123, eccentric \ac{nr} waveforms provide statistically acceptable fits --- with whitened residuals consistent with detector noise --- while preferring solutions with larger remnant masses and initial eccentricities $e_0 \gtrsim 0.5$. 
However, \nrsur{} achieves higher maximum-likelihood, indicating that while eccentric configurations are viable, they are not favoured by the data.

The remainder of this paper is organised as follows. In Section~\ref{sec:confusing}, we quantify the conditions under which eccentric signals can be misidentified as quasi-spherical. In Section~\ref{sec:incorrect}, we study the resulting biases in inferred source parameters. In Section~\ref{sec:misinterpreted}, we analyse GW231123 data using eccentric NR waveforms and assess their consistency with the observations. We summarise our findings and discuss their implications in Section~\ref{sec:conclusion}.

\section{Confusing eccentric and non-eccentric signals}
\label{sec:confusing}

\subsection{Faithfulness and model indistinguishability}
\label{sec:faithfulness}

To assess whether a \ac{gw} signal is better described by an eccentric or a quasi-spherical waveform model, one needs to compare different waveform hypotheses with the data \(d\). For a given model \(M\), parameter estimation aims to identify waveform parameters \(\boldsymbol{\theta}\) such that the template \(h(\boldsymbol{\theta}\mid M)\) provides an adequate representation of the signal \(s(\boldsymbol{\lambda})\) buried in the noisy data \(d = s(\boldsymbol{\lambda}) + n\) ~\citep{Umstatter:2007xfq, Veitch:2008ur}. 
In other words, the residual \(r = d - h(\boldsymbol{\theta}\mid M) \) must be statistically consistent with the postulated noise distribution $n$.

When the waveform hypothesis is incorrect—for example, when a quasi-spherical model is used to analyse an eccentric signal—the disagreement between the signal and the best-fitting template introduces a discrepancy \(\delta h = s - h\). Such differences can bias the inferred parameters or even render the model incompatible with the data. 
Their impact, however, can remain negligible and give rise to indistinguishable interpretations when the discrepancy satisfies \((\delta h \mid \delta h) < \epsilon\), where \(\epsilon\) is a tolerance set by the noise properties of the detector~\citep{Lindblom:2008cm, Damour:2010zb, Chatziioannou:2017tdw, Benetti:2026enj}.

A convenient measure of the statistical consequence of an incorrect signal hypothesis is the reduction in the maximum log-likelihood \(\Delta \ln \mathcal{L} \equiv \ln \hat{\mathcal{L}}(\boldsymbol{\lambda}) - \ln \mathcal{L}_{\mathrm{max}}(\boldsymbol{\theta})\) relative to the model~\footnote{Throughout this work, \(\ln \mathcal{L}\) or ``log-likelihood'' is used as shorthand for the log-likelihood ratio between the signal and noise hypotheses.}. Here, \(\ln \hat{\mathcal{L}}(\boldsymbol{\lambda})\) is the actual value of the log-likelihood at the true parameters \(\boldsymbol{\lambda}\) and \(\ln \mathcal{L}_{\mathrm{max}}(\boldsymbol{\theta})\) is the maximum log-likelihood value that the waveform model can achieve. 
This reduction is related to the faithfulness~\citep{Apostolatos:1995pj},
\begin{equation}\label{eq:fitting-factor}
    \mathcal{F}
    = \max_{\boldsymbol{\Lambda}}
    \frac{ \left( s \mid h(\boldsymbol{\theta}) \right) }
         { \sqrt{ \left( s \mid s \right) \left( h \mid h \right) } }\,,
\end{equation}
quantifying the agreement between two waveforms after maximisation over a chosen set of parameters \(\boldsymbol{\Lambda} \subset \boldsymbol{\theta}\). 
Their relationship reads
\begin{equation}\label{eq:criterion-2}
    2\Delta \ln \mathcal{L}
    \simeq \rho^2 \bar{\mathcal{F}}
\end{equation}

where \(\rho\) is the optimal \ac{snr} of the signal, \(\bar{\mathcal{F}}=1-\mathcal{F}\) is the unfaithfulness or the disagreement between the signal and waveform model.
Now, if we assume that the inferred posteriors are well-approximated by a multivariate Gaussian in \(N\) parameters and that the data can be modelled as a stationary, Gaussian stochastic process, then \(2\Delta \ln \mathcal{L}\) follows a \(\chi^2_N\) distribution~\citep{Toubiana:2024car, Thompson:2025hhc}.
Consequently, the region containing $90\%$ of the posterior volume satisfies: 
\begin{equation}\label{eq:criterion-2}
    2\Delta \ln \mathcal{L}
    \simeq \rho^2 \bar{\mathcal{F}}
    < \mathcal{Q}_{0.9}(N) \implies \bar{\mathcal{F}} < \frac{\mathcal{Q}_{0.9}(N)}{\rho^2}
\end{equation}

where \(\mathcal{Q}_{0.9}(N)\) denotes the \(90\%\) quantile of the \(\chi^2_N\) distribution. 
In the following, before turning to full parameter estimation, we will use this metric to quantify the level of agreement between the models we employ and the numerical waveforms.

\subsection{Method and dataset}
\label{sec:methods}

\begin{figure}
    \centering
    \includegraphics[width=0.48\textwidth]{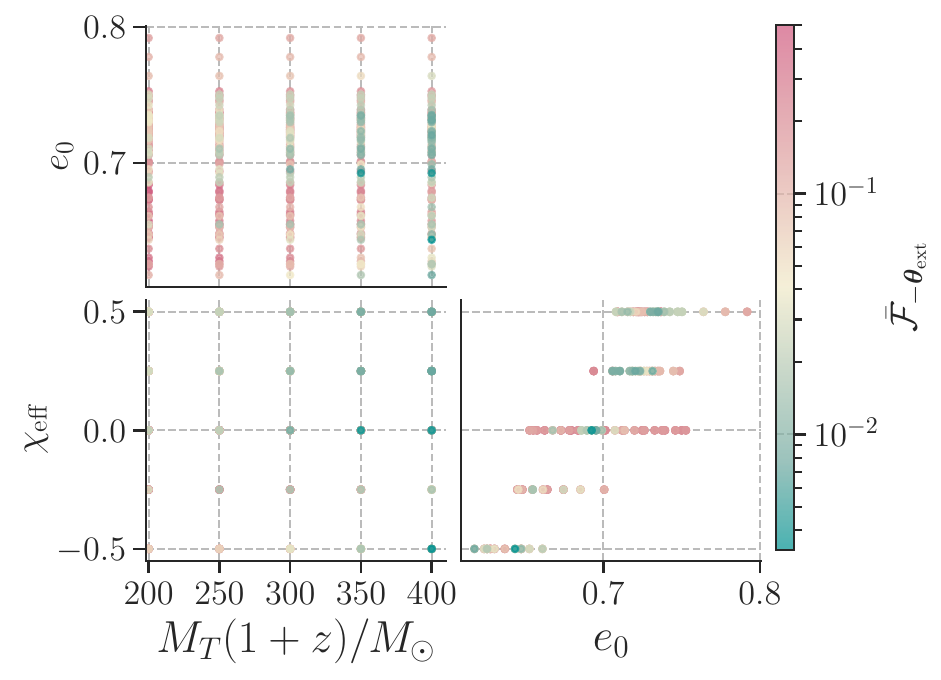}   
    \caption{Unfaithfulness \(\bar{\mathcal{F}}_{-\boldsymbol{\theta}_{\rm ext}}\) between eccentric \ac{nr} signals from the ICCUB catalogue and quasi-spherical \nrsur{} templates, computed while keeping the shared intrinsic parameters fixed between the signal and template and optimising only over the extrinsic parameters \(\boldsymbol{\theta}_{\rm ext}\). This corresponds to a waveform-modelling perspective that isolates discrepancies arising from physical effects absent in the non-eccentric waveform model. Each point denotes an eccentric \ac{nr} configuration, coloured by its corresponding unfaithfulness value. We note that some points overlap across multiple configurations. The distributions are shown as a function of detector-frame total mass, eccentricity, and effective inspiral spin \(\chi_\mathrm{eff}\). Lower values of \(\bar{\mathcal{F}}_{-\boldsymbol{\theta}_{\rm ext}}\) indicate eccentric signals that are more readily mimicked by quasi-spherical waveforms despite differences in the underlying binary dynamics.}
    \label{fig:unfaithfulness}
\end{figure}

To assess when eccentric signals can be misidentified as quasi-circular in high-mass \ac{bbh} systems, we construct $\sim\!1000$ signals by rescaling $128$ eccentric \ac{nr} waveforms from the ICCUB catalogue to detector-frame total masses in the range $200$--$400\,M_\odot$~\citep{Trenado:2025ccf}. Throughout, we construct our \ac{nr} waveforms with radiation multipoles \((\ell,m)= \{(2, \pm2), (2\pm1), (2,0), (3\pm3), (3,\pm2), (4,\pm4), (4\pm3)\}\), thereby neglecting subdominant modes that will contribute negligibly. For each signal, we compute the extrinsic-parameter-maximised faithfulness $\mathcal{F}_{-\boldsymbol{\theta}_{\rm ext}}$ defined in Eq.~\eqref{eq:fitting-factor}, where the extrinsic parameters include the signal-template's relative time offset, luminosity distance, sky location, polarisation angle, and binary orientation angles. Specifically, we use Eq.~13 of \citet{Harry:2017weg}, (see also~\citep{Chandra:2022ixv}) which analytically maximises over the sky-position angles, polarization angle, and luminosity distance. We evaluate this quantity as a function of relative time shift and numerically maximise it over the binary-orientation angles $(\iota,\varphi)$, retaining the minimum unfaithfulness over the resulting time series. Throughout, all intrinsic binary parameters shared by the signal and template are held fixed. This choice isolates discrepancies arising from physical effects absent in the waveform model, so the resulting quantity should be interpreted as a waveform-modelling measure of the signal-template discrepancy.

\subsection{Results: when eccentric signals appear non-eccentric}
\label{sec:nr-faithfulness}

Figure~\ref{fig:unfaithfulness} summarises our results. Each point corresponds to a distinct binary, coloured by $\bar{\mathcal{F}}_{-\boldsymbol{\theta}_{\rm ext}}$, with projections onto detector-frame total mass $M_T(1+z)$, initial eccentricity $e_0$, and $\chi_{\rm eff}$~\citep{Santamaria:2010yb}. We note that when projected onto two dimensions, the points overlap. The initial eccentricity, defined for the RIT and ICCUB catalogues as the eccentricity parameter used to construct the initial data at apocentre, is obtained directly from the simulation metadata~\citep{Healy:2022wdn, Trenado:2025ccf}. In these catalogues, the tangential momentum is reduced relative to its quasi-circular value according to $p_t=p_{t,\rm qc}(1-\epsilon)$, yielding an approximate initial eccentricity $e_0\simeq2\epsilon-\epsilon^2$. The SXS catalogue adopts a different definition. Rather than specifying the eccentricity through the initial-data construction, SXS estimates eccentricity from the orbital trajectories using post-Newtonian-inspired fits to the orbital motion, with different fitting procedures employed depending on the eccentricity regime~\citep{Scheel:2025jct}. We use these catalogue-reported eccentricities rather than deriving a common measure from the waveforms or orbital trajectories by hybridising with semi-analytical waveforms.

The dominant trend is the expected monotonic decrease in $\bar{\mathcal{F}}_{-\boldsymbol{\theta}_{\rm ext}}$ with increasing total mass, indicating that it becomes progressively difficult to distinguish an eccentric signal from non-eccentric ones. As $M_T(1+z)$ grows, inspiral cycles are progressively redshifted out of the detector band, and the observable signal becomes dominated by the remnant BH ringdown. 
The ringdown signal is dominated by the remnant mass and spin, which are determined predominantly by the progenitor masses and spins, with only a subdominant contribution from eccentricity and mean anomaly~\citep{Carullo:2023kvj, Khairnar:2024rzs}. 
Consequently, \(\bar{\mathcal{F}}_{-\boldsymbol{\theta}_{\rm ext}}\) falls, allowing for multiple, comparably likely interpretations~\citep{Chandra:2023nge, Miller:2025eak}. 
Although ringdown amplitudes retain imprints of the progenitors~\citep{Carullo:2024smg, Rao:2026lmz}--- since the perturber's properties influence the excitation of ringdown modes~\citep{DeAmicis:2025xuh, Kuchler:2025hwx, Berti:2025hly}—these features are expected to be comparatively difficult to observe at current \ac{snr}.

At fixed total mass, increasing $\chi_{\rm eff}$ generally leads to smaller unfaithfulness values, although this trend breaks down for \(M_T \geq 350 M_\odot\). 
The dependence on $e_0$, by contrast, is non-monotonic at any total mass, following as expected the trend of dominant waveform quantities such as peak amplitude~\citep{Carullo:2023kvj}. Taken together, these effects corroborate previous findings~\citep{CalderonBustillo:2020xms, Romero-Shaw:2022fbf, Divyajyoti:2025cwq}. When the above analysis is repeated with eccentric waveforms from the SXS and RIT catalogues~\citep{Scheel:2025jct, Healy:2022wdn}, broadly similar trends are observed. 

We emphasise that the quantity \(\bar{\mathcal{F}}_{-\boldsymbol{\theta}_{\rm ext}}\) isolates waveform-modelling discrepancies by keeping the intrinsic binary parameters fixed between the signal and template. In a true Bayesian inference analysis, however, both intrinsic and extrinsic parameters are allowed to vary simultaneously, enabling waveform differences between eccentric and quasi-spherical signals to be partially absorbed via adjustments to intrinsic parameters. To assess the impact of this effect, we additionally compute the fully optimised unfaithfulness \(\bar{\mathcal{F}}\), maximising over the complete parameter space using a combination of differential evolution and the Limited-memory BFGS algorithm~\citep{Storn:1997uea, RichardH:2006hnx}. We find qualitatively similar behaviour and comparable order-of-magnitude values, although the resulting unfaithfulness is systematically lower than in the extrinsic-only case, as expected.

To translate these unfaithfulness values into a statement about observational consequences, we use Eq.~\eqref{eq:criterion-2} with $N=8$ degrees of freedom, which yields $\mathcal{Q}_{0.9}(8)\approx 13.4$. For $M_T(1+z)\gtrsim 300\,M_\odot$, we have
$\bar{\mathcal{F}}_{-\boldsymbol{\theta}_{\rm ext}}\in(0.003,\,0.03)$, which implies that the eccentric and quasi-spherical families become statistically distinguishable only at $\rho \in (21,\,67)$. However, when viewed through the lens of the Akaike information criterion (AIC)~\citep{Akaike:1998zah}, with the eccentric model carrying two additional parameters ($e_0$ and mean anomaly $l_0$), we obtain $\Delta\mathrm{AIC}^{\mathrm{non\text{-}ecc}}_{\mathrm{ecc}} = \rho^2\bar{\mathcal{F}} - 4 \in (-2.8,\,8.0)$ for $\rho=20$, indicating that even for the most massive systems, the quasi-spherical model is only moderately disfavoured.
The penalty for ignoring eccentricity and mean anomaly is smaller than the cost of the extra parameters.

\section{Astrophysical consequences of incorrect modeling}
\label{sec:incorrect}

\begin{table}
\centering
\caption{True parameter values common to all simulations in Sec.~\ref{sec:incorrect}. These values are consistent with the binary properties of GW231123 as inferred by the \ac{lvk} using the \nrsur{} waveform model. All angular quantities are given in radians.}
\begin{tabular}{|lll|}
\hline Parameter & Symbol & Value \\
\hline \hline Detector-frame total mass & $M_{T}(1+z)$ & $320 M_{\odot}$ \\
Inclination angle & \(\iota\) & 0.77 \\
Azimuth & $\phi$ & 1.22 \\
Polarization angle & $\psi$ & 1.29 \\
Geocentric peak time & $t_{\rm gps}$ & 1384782888.7s \\
Right ascension & $\alpha$ & 2.39 \\
Declination & $\delta$ & 0.29 \\
\hline
\end{tabular}
\label{table:properties-1}
\end{table}

The previous section establishes that, for sufficiently high-mass systems, an eccentric waveform can be well reproduced by a quasi-spherical template, raising the prospect of systematic bias if the wrong source model is assumed. We now quantify this bias directly using two different approaches.

We select the subset of eccentric \ac{nr} waveforms that yield the smallest unfaithfulness $\bar{\mathcal{F}}$ against the best-fitting \nrsur{} waveform for GW231123. 
We remark this is a conservative choice, and for higher unfaithfulness we expect higher biases. 
We set all common extrinsic parameters 
to the values in Table~\ref{table:properties-1}, set the luminosity distance $D_L$ to give a network optimal \ac{snr} of $22$ (consistent with GW231123) and analyse the signals in zero noise with a two-detector network comprising Advanced LIGO detectors operating with a power spectral density comparable to that at the time of GW231123.

We assess the induced bias through two complementary approaches:
\begin{inparaenum}[(1)]
    \item a full frequency-domain Bayesian inference of the entire signal with a quasi-spherical \ac{bbh} waveform model, which tests the global consistency of the waveform under the assumption of a non-eccentric progenitor; and
    \item a targeted time-domain post-peak ringdown analysis using multiple Kerr quasinormal modes combinations, including only data after different times relative to the waveform peak, which probes the remnant properties.
\end{inparaenum}

These two analyses are complementary because they are sensitive to different aspects of the signal. The full inference is susceptible to any eccentricity-induced phase and amplitude modulations accumulated over the entire waveform and to the lack of noncircular calibration in the merger-ringdown sector.
The post-peak analysis isolates the regime where the remnant is already formed, and the signal is well described by damped sinusoids. Since the ringdown spectrum is governed by the remnant mass and spin (see discussion above), we expect the post-peak analysis to be robust to the orbital configuration, even when the full inference is not. 
Our focus is therefore not on the detectability of eccentric imprints per se, but on how the modelling assumption propagates into parameter biases and whether the post-peak remnant characterisation remains reliable despite it.

\subsection{Full-signal analyses}

\begin{figure*}
    \centering
\includegraphics[width=0.8\textwidth]{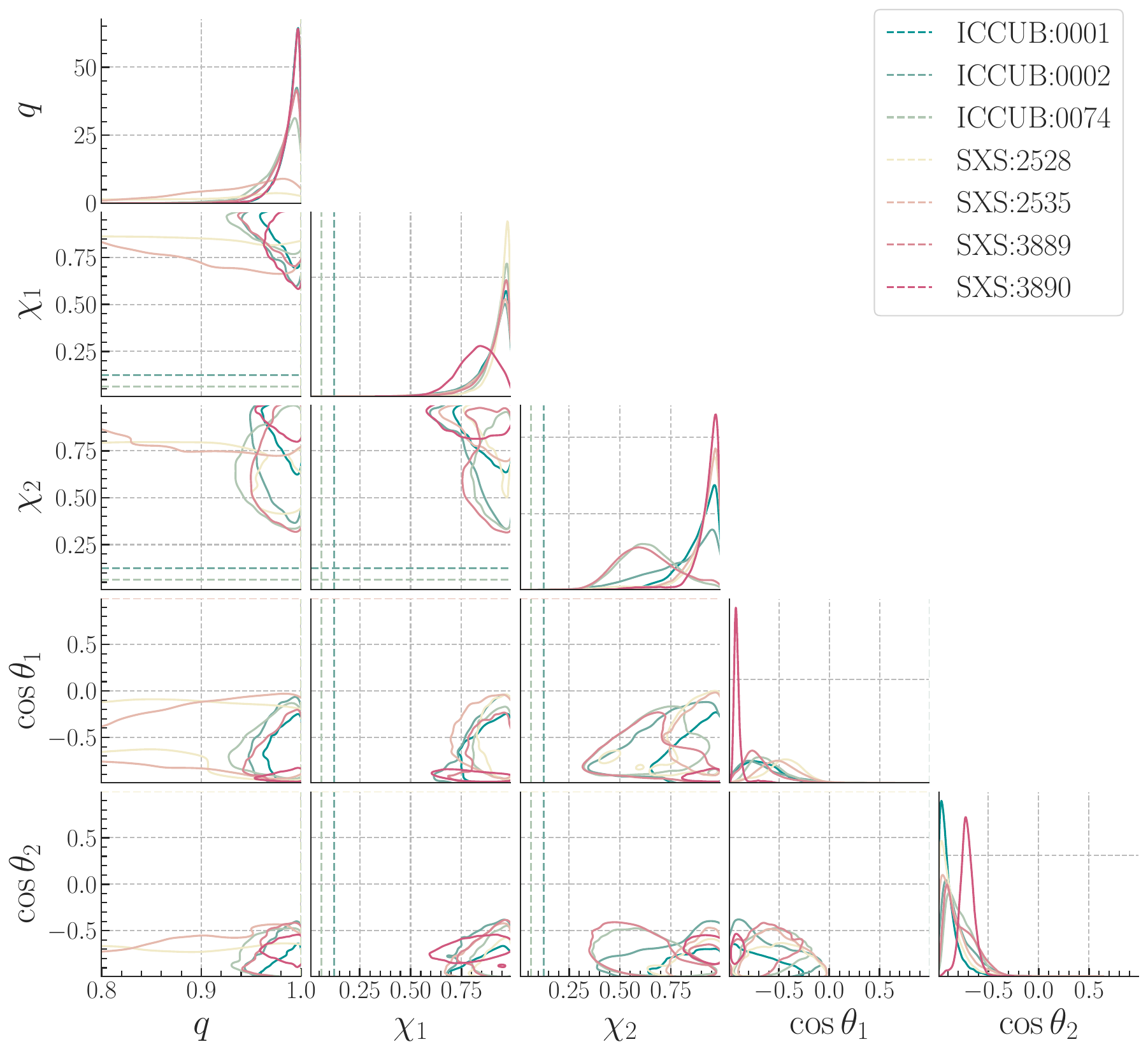}
    \caption{Inferred Posteriors of intrinsic binary parameters when eccentric signals are analysed with \nrsur{} waveforms. Each colour corresponds to a different eccentric simulation with parameters summarised in Table~\ref{table:simulation-details}. Despite the underlying source having low spins, the inferred source has at least a large primary spin, illustrating how eccentric effects can be absorbed through correlated shifts in quasi-spherical parameters. 
    Some of the injected values are not displayed due to being at the boundaries of the axes.}
    \label{fig:icc-nr}
\end{figure*}

\begin{figure}
    \centering
    \includegraphics[width=0.98\linewidth]{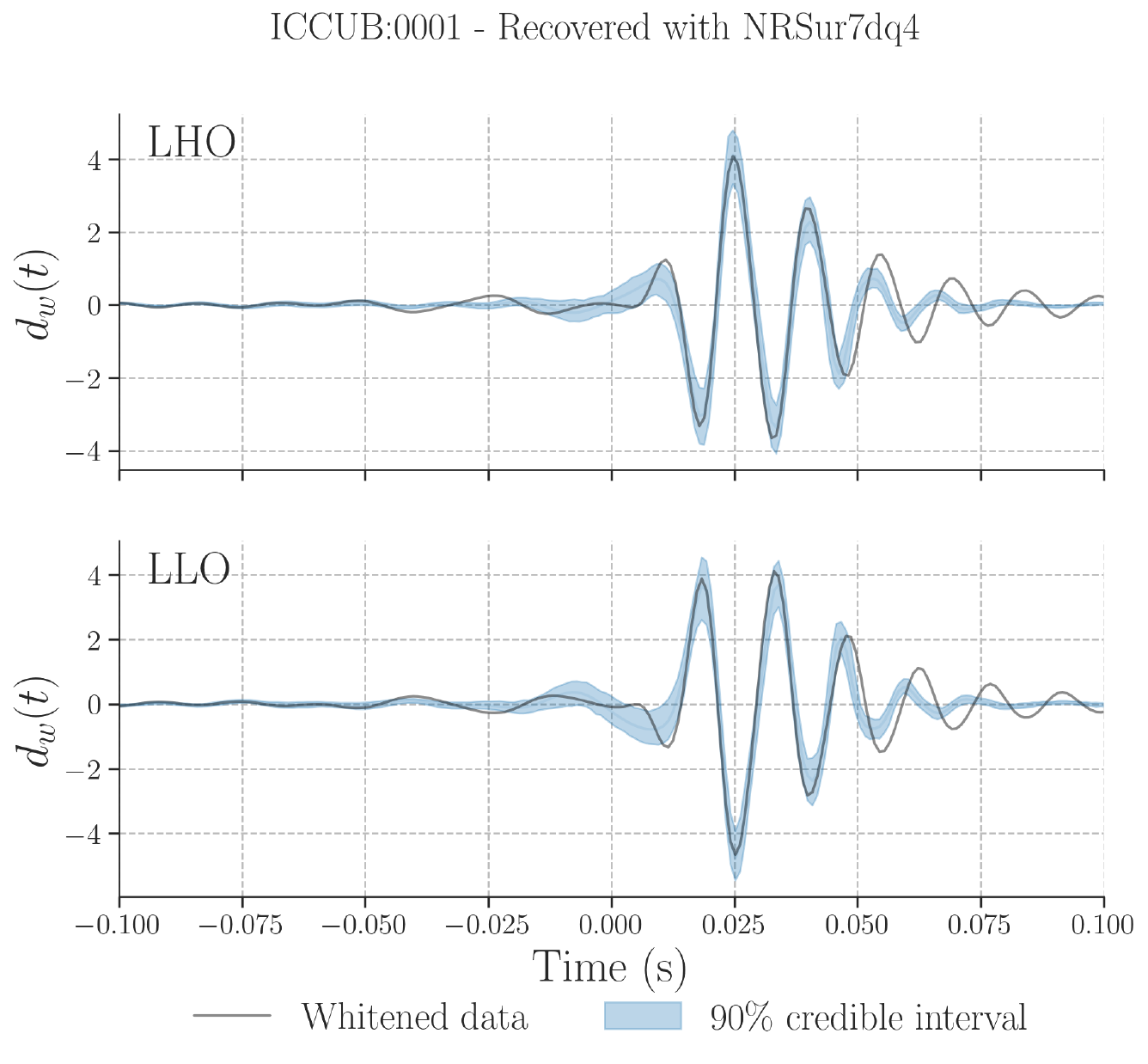}
    \caption{Whitened time-domain strain in the Hanford (top) and Livingston (bottom) detectors. The \ac{nr} simulation ICCUB:0001 is analysed using the quasi-spherical \nrsur{} waveform model in zero noise. Despite providing a reasonable fit around merger, the recovered waveform fails to reproduce the post-peak structure of the \ac{nr} signal.}
    \label{fig:icc-recovery}
\end{figure}

\begin{table}
    \centering
    \begin{tabular}{|lllllll|}
    \hline Simulation ID  & \(q\) & \(e_0\) & \(\chi_1\) & \(\chi_2\) & \(e_{gw}^{5 \rm Hz}\) & \(\ell_{gw}^{5 \rm Hz}\)\\
    \hline \hline ICCUB:0001 & 1 & 0.79 & 0.13 & 0.13 & - & -\\
    ICCUB:0002 & 1 & 0.78 & 0.13 & 0.13 & - & - \\
    ICCUB:0074 & 1 & 0.73 & 0.06 & 0.06 & - & - \\
    \hline
    SXS:2535 & 0.5  & 0.58 & 0. & 0. & 0.40 & 6.16 \\
    SXS:2528 & 1.   & 0.62 & 0. & 0. & 0.40 & 1.75 \\ 
    SXS:3889 & 1.   & \(> 1\)  & 0. & 0. & - & - \\
    SXS:3890 & 0.67 & \(> 1\)  & 0. & 0. & - & - \\    
    \hline
\end{tabular}
    \caption{
    The identifiers and key physical properties of the eccentric \ac{nr} simulations used in this analysis. 
    With $e_0 >1$ we indicate dynamically bound orbits, corresponding to hyperbolic orbits in Newtonian gravity. 
    Throughout the paper, we define the mass ratio \(q = m_2/m_1 \leq 1\).
    }
\label{table:simulation-details}
\end{table}

Our full-signal analysis results are shown in Figure~\ref{fig:icc-nr}, when the simulated signal consists of eccentric \ac{nr} waveforms from the ICCUB and SXS catalogue~\citep{Trenado:2025ccf, Scheel:2025jct}.
When recovered with the quasi-spherical waveform model, a clear preference for large dimensionless spins \(\chi_k \gtrsim 0.8\) at 90\% credibility is found, at least for the more massive \ac{bh}, despite the simulated waveforms corresponding to systems with low component spins. 
While the waveforms in the first catalogue correspond exclusively to elliptic binaries, those in the second catalogue include a mixture of elliptic and dynamically bound configurations, the latter marked by \(e_0 > 1\) since they would correspond to hyperbolic trajectories in Newtonian gravity. 
The observed behaviour is consistent with the optimisation-based picture discussed in Sec.~\ref{sec:faithfulness}: eccentric features in the signal can be partially absorbed by correlated shifts in the quasi-spherical parameter space.
Although the detailed features of the posteriors vary across catalogues, reflecting differences in eccentricity definitions, waveform construction, and numerical systematics, the qualitative conclusion remains unchanged: for short-lived, high-mass signals, quasi-spherical waveform models can incorrectly infer high spin components, especially when the true signal is highly eccentric. In particular, as shown in Figure~\ref{fig:icc-recovery} for ICCUB:0001, the reconstructed \nrsur{} waveforms across all simulations fail to adequately capture the post-merger signal.

Viewed from an information-criteria perspective, and assuming the existence of an eccentric precessing waveform model capable of accurately fitting the signal, we find that the quasi-spherical hypothesis is strongly disfavoured even at this moderate network \ac{snr}. For these representative signals, the sampler at best recovers \(\rho \simeq 21\); this evaluates to \(\Delta \mathrm{AIC}^{\mathrm{non\text{-}ecc}}_{\mathrm{ecc}}\gtrsim 39\), thereby decisively favouring the eccentric hypothesis based on the data alone.

At higher \ac{snr}, the increased statistical power further amplifies this preference, and the likelihood gain from using a correct waveform model will outweigh the Ockham penalty associated with the prior space. These results demonstrate that, while quasi-spherical waveform models can produce ``apparently'' well-constrained and astrophysically plausible posteriors, model selection, even when neglecting prior space, can decisively disfavour them.

\subsection{Post-peak analyses}

At sufficiently late times, the relaxation phase following the merger is well described by a superposition of exponentially damped sinusoids—the \acp{qnm} of the remnant BH. Restricting to the corotating modes, which dominate over their counter-rotating counterparts~\citep{JimenezForteza:2020cve, Dhani:2020nik, Li:2021wgz}, the Kerr template for the complex strain at the detectors, within the stationary \ac{qnm} regime, takes the form
\begin{equation}\label{eq:kerr}
    h = \sum_{\substack{\ell\geq2\\0\leq m\leq\ell\\n\geq0}}
    A_{\ell mn}\,e^{-i\omega_{\ell mn}(t-t_{\rm start}) + i\phi_{\ell mn}}\,
    {}_{-2}S_{\ell mn}(M_{f} \, \chi_{f} \, \omega_{\ell mn},\iota,\varphi)~.
\end{equation}
Here $\omega_{\ell mn} = 2\pi f_{\ell mn} - i/\tau_{\ell mn}$ is the complex frequency, with $(\ell,m,n)$ labelling the angular and overtone indices, and ${}_{-2}S_{\ell mn}$ the spin weighted spheroidal harmonics.
Within \ac{gr}, for astrophysical Kerr BHs $\omega_{\ell mn}$ depends only on the detector-frame remnant mass $M_f(1+z)$ and dimensionless spin $\chi_f$~\citep{Press:1971wr,Chandrasekhar:1975zza,Detweiler:1980gk,Dreyer:2003bv,Berti:2005ys,Berti:2009kk}; the complex amplitudes $A_{\ell mn}$, inclination \(\iota\) and phases $\phi_{\ell mn}$ are left as free parameters alongside $M_f(1+z)$ and $\chi_f$.

\begin{figure}[b]
    \centering
    \includegraphics[width=0.48\textwidth]{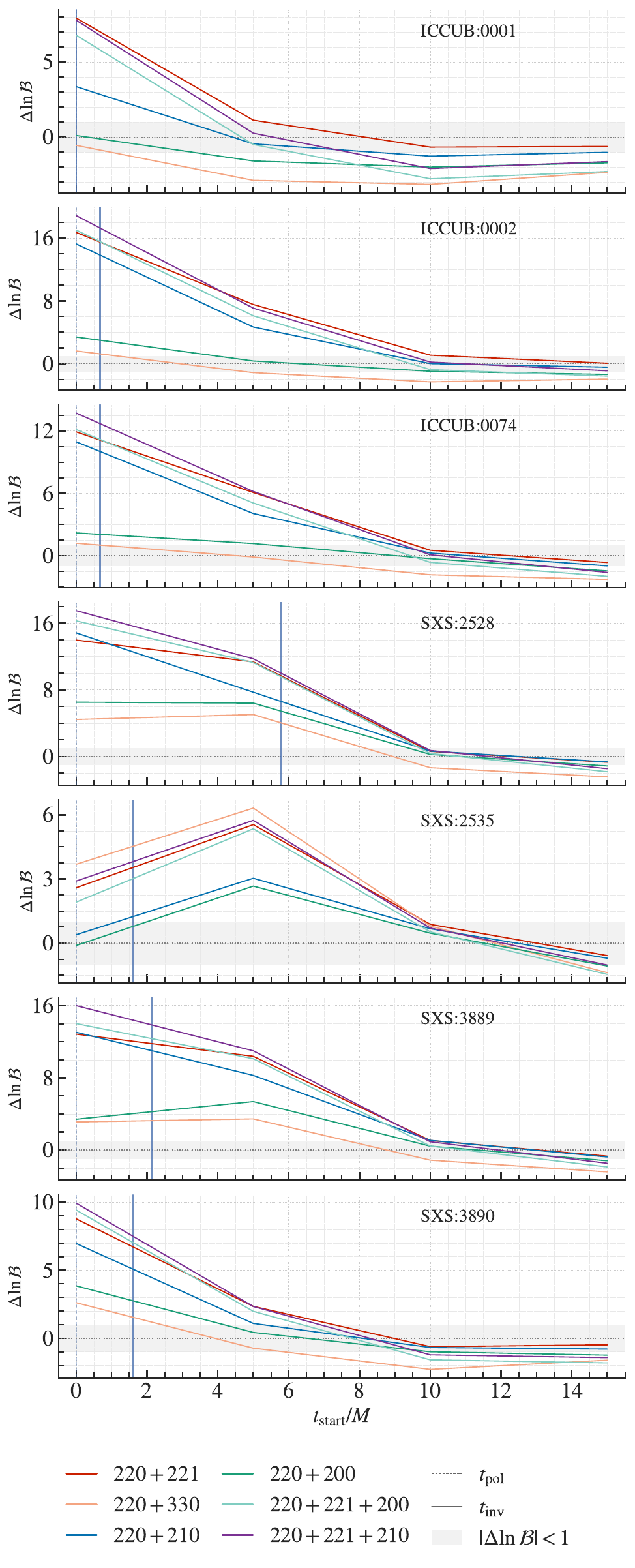}
    \caption{The Bayes factors comparing the analysed Kerr model to the nested fundamental mode-only model, with the time axis relative to the polarisation peak.}
    \label{fig:bayes-factor-kerr}
\end{figure}

We use \texttt{pyRing}~\citep{Carullo:2019flw,pyRing} to fit the Kerr template from Eq.~\eqref{eq:kerr} to the signals listed in Table~\ref{table:simulation-details}.
To avoid contamination from data that are not described by Eq.~\eqref{eq:kerr}, we excise data before a given \(t_{\rm start}\).
Given the uncertainty in the onset of the Kerr-dominated regime, the analysis is repeated for a range of starting times \(t_{\rm start}\) relative to the waveform polarisation peak, \(t_{\mathrm{peak}}^{\mathrm{pol}}=\max_t |h_+ - i h_\times|^2 \). This is different from the invariant strain peak defined as \(t_{\rm inv} = \max_t |h_{\ell,m}|^2\). We define them using the \ac{nr} simulations.
For the recovery, we consider several $( \ell,m,n)$ mode combinations in addition to the dominant \(220\) mode, comprising the ones expected to dominate binary merger relaxation: $\{ (221), (210), (200), (330) \}$.

Figure~\ref{fig:bayes-factor-kerr} compares the natural log Bayes Factor (\(\ln \mathcal{B}\)) of the different Kerr models with respect to \(\mathrm{Kerr}_{220}\) as a function of \(t_{\rm start}\) measured in units \(t_M=GM_f^{\rm NR}(1+z)/c^3 \). 
All simulations show an overall similar trend, namely overwhelmingly positive evidence for more than a single mode at early times, rapidly decaying away from the peak.
The flexibility of the Kerr template means that at times close to the peak, additional modes can act as a flexible basis to absorb non-\ac{qnm}, non-stationary \ac{qnm} and non-linear components present at early times~\citep{DeAmicis:2025xuh, Berti:2025hly}, artificially inflating the evidence without reflecting genuine \ac{qnm} excitation. 
Instead, at times later than \(t_{\rm start} \gtrsim 10 t_M\), the template is expected to become a sufficiently accurate approximation, for SNRs comparable to GW231123.

In this regime, we find that the preferred modes are typically the $\{ (221), (210) \}$, in agreement with theoretical expectations~\citep{Cheung:2023vki}.
Still, the evidence for these additional modes never reaches statistical significance, indicating that the contributions beyond the fundamental mode are negligible and that adding them to the model incurs an Ockham penalty without improving the fit.
Note that for the five equal-mass equal-spin simulated configurations, odd-$m$ modes are identically zero due to exchange symmetry.
We find that the SXS simulations with the exception of SXS:3890 show a kink at $5t_M$ the reason for which is unknown. We hope to explore this in a future work.

\begin{figure*}
    \centering
    \includegraphics[width=0.98\textwidth]{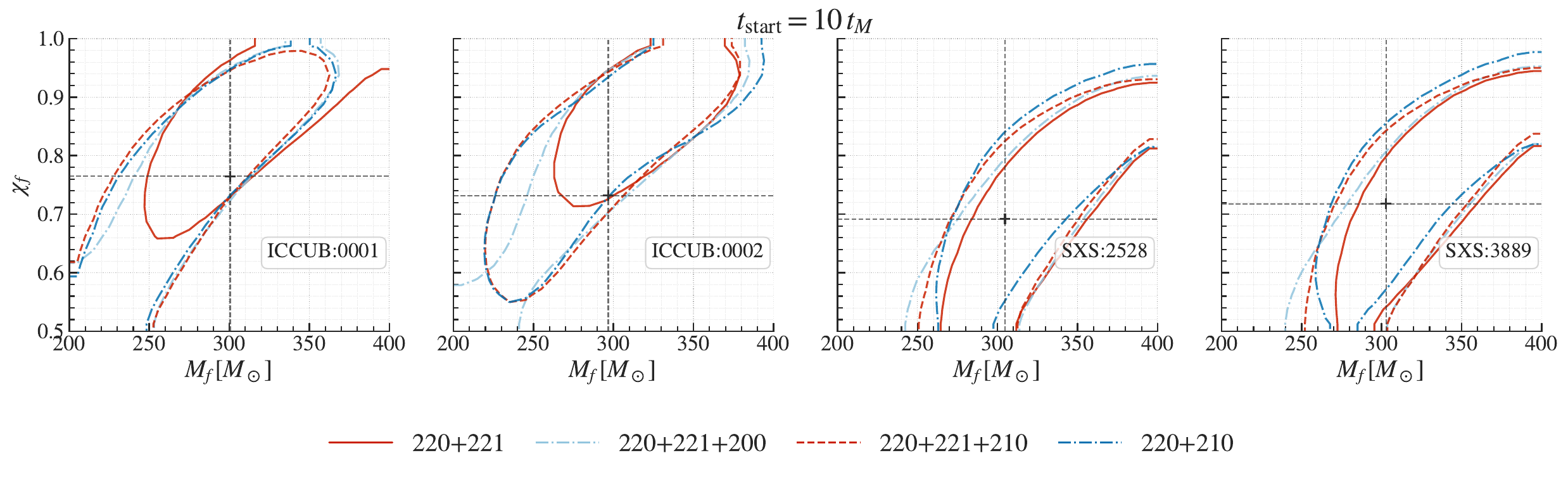}
    \caption{Mass and spins of the remnant \ac{bh} for different eccentric \ac{nr} waveforms, starting at their nominal validity time, \(t_{\rm start}=10t_M\). The contours enclose 90\% of the posterior samples, and the dashed black lines indicate the true values.}
    \label{fig:final-mass-spin-kerr}
\end{figure*}

In Figure~\ref{fig:final-mass-spin-kerr} we show the remnant mass and spin estimated using selected Kerr models at the nominal start of their validity regime \(t_{\rm start}=10t_M\), for representative simulations from each catalogue. 
The posteriors obtained using the different Kerr models recover the true remnant parameters within the 90\% credible regions for both catalogues, confirming the robustness of the remnant inference. 
For the ICCUB, the true values lie at the edges of the 90\% credible regions, particularly for ICCUB:0002, while for the SXS, recovered parameters are consistent with the simulated values.

\section{Is GW231123 misinterpreted?}
\label{sec:misinterpreted}

\begin{figure}[t!]
    \centering
    \includegraphics[width=0.5\textwidth]{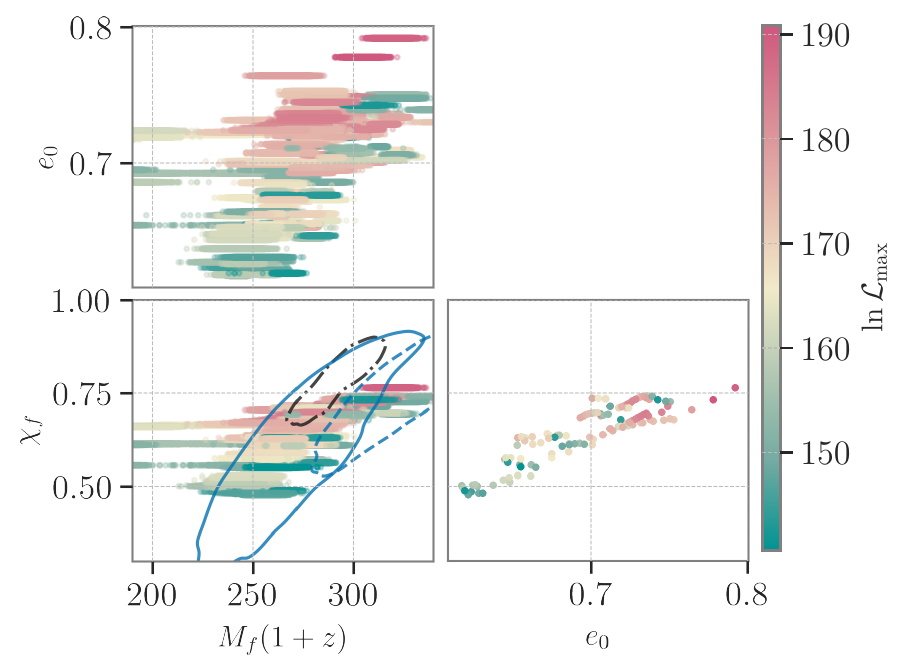}
\caption{
Redshifted final mass \(M_f(1+z)\), final spin \(\chi_f\), and initial eccentricity \(e_0\) for GW231123 inferred from different waveform models. The black contour in the lower-left panel encloses the 90\% credible region obtained with the \nrsur{} waveform model, the blue dashed (solid)  contour encloses the 90\% credible interval of a ringdown-only analysis performed using Kerr\(_{220}\) template at 12\(t_M\) (20\(t_M\)) while the scattered points in all the subpanels correspond to \ac{nr}-based inferences, coloured according to their maximum (over extrinsic parameters) log-likelihood $\ln \mathcal{L}$.}
\label{fig:final-mass-spin}
\end{figure}

We next analyse 8\,s of publicly available Advanced LIGO detector data containing GW231123, downsampled to \(1024\,\mathrm{Hz}\), using our suite of eccentric \ac{nr} simulations and the parameter-estimation package \bilby. 
Our suite consists of \ac{nr} simulations from the ICCUB, RIT, and SXS catalogues~\citep{Trenado:2025ccf, Healy:2022wdn, Scheel:2025jct} whose initial eccentricity \(e_0 \gtrsim 0.5\), as defined by the respective catalogue conventions, and with high-quality waveforms.
Consistent with the \ac{lvk} analyses, we employ a minimum and maximum frequency cutoff of 20\,Hz and 448\,Hz, respectively, whilst stochastically evaluating the Whittle Likelihood~\citep{whittle_estimation_1953}.
For the extrinsic parameters, we use priors consistent with \ac{lvk}'s analyses, while for \(M_T(1+z)\), we use a flat prior over \(100-400 M_\odot\).

The left--bottom panel of Fig.~\ref{fig:final-mass-spin} compares the joint posteriors for the redshifted remnant mass \(M_f(1+z)\) and remnant dimensionless spin \(\chi_f\) inferred using eccentric \ac{nr} waveforms with those obtained by the \ac{lvk} using the non-eccentric \nrsur{} model (black dash--dotted contours). 
Each point corresponds to an independent \ac{nr}-based inference from the ICCUB catalogue and is coloured by the corresponding maximum log-likelihood $\ln \mathcal{L}_{\rm max}$.
While some eccentric-\ac{nr} posteriors overlap with the \nrsur{}'s 90\% credible region, the simulations giving the highest likelihood values are systematically displaced from it, tending toward higher inferred remnant masses and spins. 
We find that the higher the inferred final mass, the higher the inferred initial eccentricity \(e_0\), as shown in the associated panels, with higher-likelihood solutions preferentially occurring at larger \(e_0\). 
This result highlights how waveform-model differences can propagate into systematic differences in estimates of remnant mass and spin.

\begin{figure}
    \centering
    \includegraphics[width=0.48\textwidth]{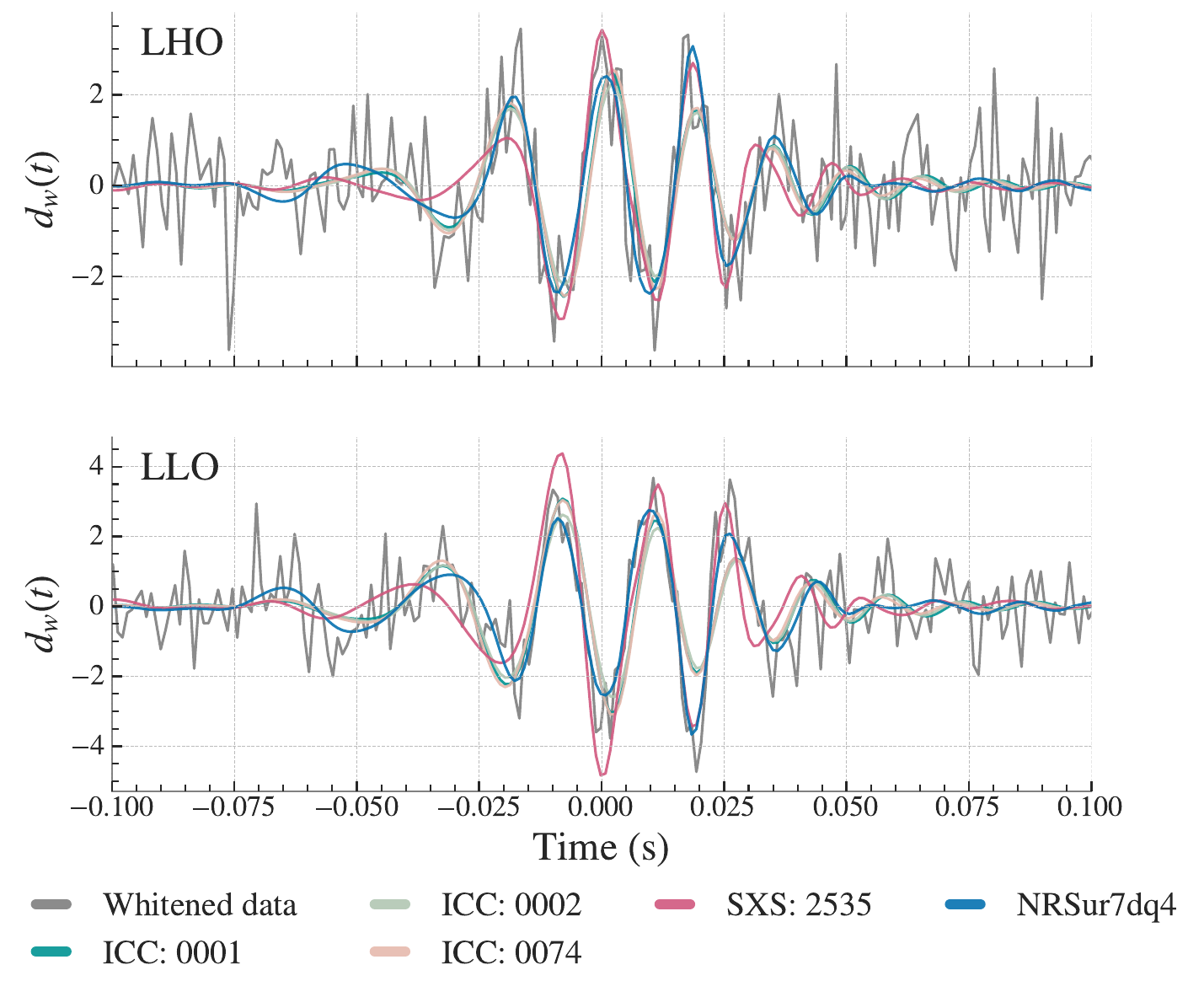}
    \caption{Whitened time‑domain data for the Hanford (top) and Livingston (bottom) detectors compared with best‑fitting \nrsur{} and \ac{nr} waveforms. Differences in how the models fit the data near the signal’s whitened strain peak dominate the likelihood and favour the \nrsur{} reconstruction.}
    \label{fig:best}
\end{figure}

Figure~\ref{fig:best} shows best-fitting \nrsur{} and representative \ac{nr} whitened waveforms recovered from our analyses, displayed in the time domain (left) and the frequency domain (right). We find that the best-fit \nrsur{} waveform (blue) more accurately describes the whitened data morphology (grey) near the signal's whitened strain peak, especially in the Livingston detector (LLO), whereas the ICCUB waveforms better describe the data preceding and after this peak, reproducing the subtle amplitude and phase modulations. The \textsc{SXS:2535} waveform also fits the data reasonably well, but given that the peak features dominate the observed signal, the \nrsur{} analysis achieves a higher log-likelihood as compared against the \ac{nr} waveforms. 

Finally, we note that \ac{nr} waveforms from the RIT catalogue consistently yield poorer fits and thereby poorer \(\ln \mathcal{L}\) as compared to those obtained using SXS and ICCUB catalogues and are hence not shown.

\begin{figure}
    \centering
    \includegraphics[width=0.48\textwidth]{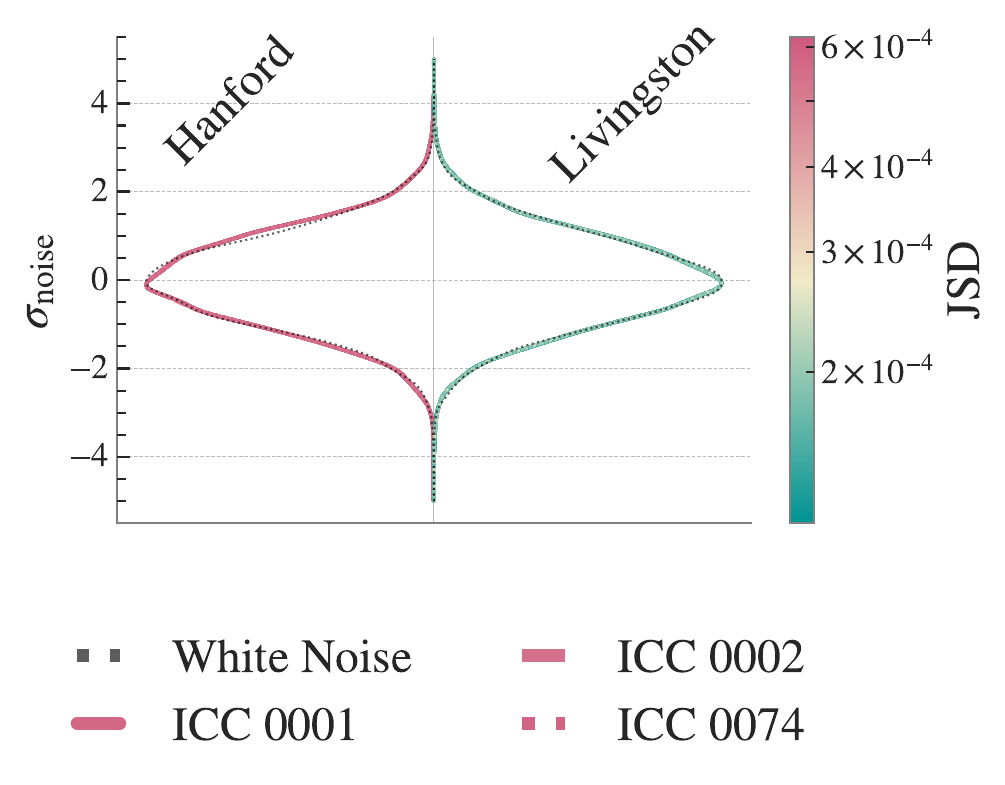}
    \caption{
    %
    %
    Time-domain residual distributions for the Hanford (left) and Livingston (right) detectors, obtained by subtracting whitened posterior-supported \ac{nr} waveforms from the whitened detector data. 
    Each curve corresponds to residuals computed from 500 posterior draws for the four best-fit \ac{nr} simulations shown in the legend, for the parameters not determined by the simulation. 
    Note that the curves overlap.
    The black dashed curve denotes a unit normal distribution, expected from an idealised PSD estimation and signal model. Simulations are coloured according to the Jensen--Shannon divergence (JSD) between their residual distribution and the noise distribution. 
    We find \(\mathrm{JSD} \lesssim 10^{-3}\) for all cases, indicating that the residuals are statistically consistent with noise and that the data are well described by the \ac{nr} waveforms.}
    \label{fig:residual}
\end{figure}

Because stochastic samplers such as \dynesty\ are designed to map out posteriors rather than locate the global maximum of the likelihood, a more faithful representation of the inference is the 90\% credible interval of the posterior-supported waveforms. Accordingly, in Figure~\ref{fig:residual}, we show violin plots of the whitened residuals, \(r_w = d_w - h_w \), constructed from posterior-supported waveforms and coloured by their Jensen--Shannon divergence (JSD) relative to a white-noise distribution (black), with line styling again indicating \ac{nr} simulation ID~\citep{Chandra:2025jfc}. We find that a subset of these posterior-supported waveforms yields whitened residuals with JSD $< 10^{-3}$, consistent with Figure~\ref{fig:best}, which shows that these eccentric \ac{nr} templates can adequately represent the observed signal within the detector band.

Owing to the fundamentally different prior volumes and parameter-space dimensionalities, we do not perform Bayesian model selection between the eccentric \ac{nr} catalogue and the \nrsur{} model. The former spans a discrete catalogue with fixed intrinsic parameters, whereas the latter is a continuous waveform model with a continuous prior. Instead, we compare the distributions of log-likelihood values under each model. We find that the \nrsur{} analysis consistently yields higher log-likelihoods, with the distribution shifted toward larger values \(\ln \mathcal{L} > 208^{+4.0}_{-5.8}\), indicating that a quasi-spherical interpretation of the GW231123 signal, rather than an elliptical \ac{bbh} signal with high eccentricity, is preferred. 
We do, however, note that the \nrsur{} analyses yield consistently high in-plane component spins \(\chi_{1,2} \gtrsim 0.9\), a region of parameter space where it is not calibrated against \ac{nr} waveforms.
Further, it yields \(\bar{\mathcal{F}}> 10^{-3}\) when compared against SXS waveforms with similarly high spins (see Figure 2 of \citet{LIGOScientific:2025rsn}), indicating its susceptibility to waveform systematics. 
Additionally, we note that the Livingston detector, where the \nrsur{} waveforms fit relatively better, also drives the spin measurements in the coherent analyses and does not agree with the spin measured with the Hanford detector-only analyses (see Appendix B of \citet{LIGOScientific:2025rsn}). 

Our analysis differs from previous results in the literature based on semi-analytical waveform models that assume effective circularisation or sphericalisation of the progenitor binary at merger~\citep{Jan:2025zcm, Lange:2026eqx, Pompili:2026yxq}.
Such analyses are intrinsically limited in this regime, as their post-merger phase descriptions rely on \ac{nr} fitting formulae calibrated to quasi-circular or quasi-spherical binaries, with mode amplitudes constructed to ensure smooth matching to a pre-merger inspiral informed by \ac{nr} simulated quasi-circular binary and test-mass signals. For GW190521- and GW231123-like signals, the inspiral contributes negligibly to the observed strain, rendering the inference particularly sensitive to these modelling assumptions. Consequently, claims favouring specific source hypotheses must be interpreted with caution when based on waveform templates whose post-merger structure implicitly encodes circularised progenitors.

\subsection{Comparison to ringdown-only analyses}

Given the possible systematic uncertainties present in both classes of \ac{imr} models, we compare their inference to more robust ringdown-only estimates.
The key subtlety in this analysis is that, to correctly interpret the ringdown measurement, a reliable estimate of the signal peak is required.
For GW231123 this parameter is subject to systematic uncertainties~\citep{LIGOScientific:2025rsn}.
When assuming $t_{\mathrm{peak}}^{\mathrm{pol}}$ predicted by the \nrsur{}, we find that an independent ringdown-only \(220\) analysis (blue dashed contours) starting at $t_{\mathrm{peak}}^{\mathrm{pol}} + 12\,t_M$ prefers higher values of $M_f(1+z)/M_\odot$ and $\chi_f$ in agreement with eccentric models, and in tension with \nrsur{}.
As simulated analyses validate the robustness of pure ringdown at these starting times, this would seem to indicate that \ac{imr} eccentric configurations constitute a more robust physical configuration for this signal.
In this case, the lower likelihood value would possibly be due to the lack of spin-precession effects.
Clearly, this interpretation remains speculative given current available data.

To go beyond, one could note that the quasi-circular \nrsur{} model is known to be missing some coherent component of the GW231123, as indicated both by ringdown analyses~\citep{LIGOScientific:2025rsn,Siegel2025} and analyses performed under the hypotheses of overlapping or lensed signals~\citep{Hu:2025lhv, Chan2025, Goyal2025, Chakraborty2025, Shan2025, Harshe2026}.
Hence, \nrsur{}-inferred peaktime is not necessarily expected to be a robust quantity to an independent ringdown analysis.
In this case, a more robust estimate of the signal peak was argued to be given by the strain peak~\citep{LIGOScientific:2025rsn}.
With respect to this new peak, an analysis at sufficiently late times is achieved (in our conventions) by starting at $t_{\mathrm{peak}}^{\mathrm{pol}} + 20\,t_M$.
The result of this analysis, shown in Fig.~\ref{fig:final-mass-spin}, prefers remnant values in agreement with \nrsur{}, and in tension with the most likely eccentric simulations.

In summary, our analysis confirms the possibility that the remnant estimates provided by the \nrsur{} might be biased towards lower masses and higher spins.
However, better \ac{imr} models or analysis methodologies will be required to ascertain the correct interpretation of GW231123 data, and the most conservative description remains in agreement with \nrsur{}.

\subsection{On the exceptionality of the GW231123 signal}

\begin{figure}
    \centering
    \includegraphics[width=0.46\textwidth]{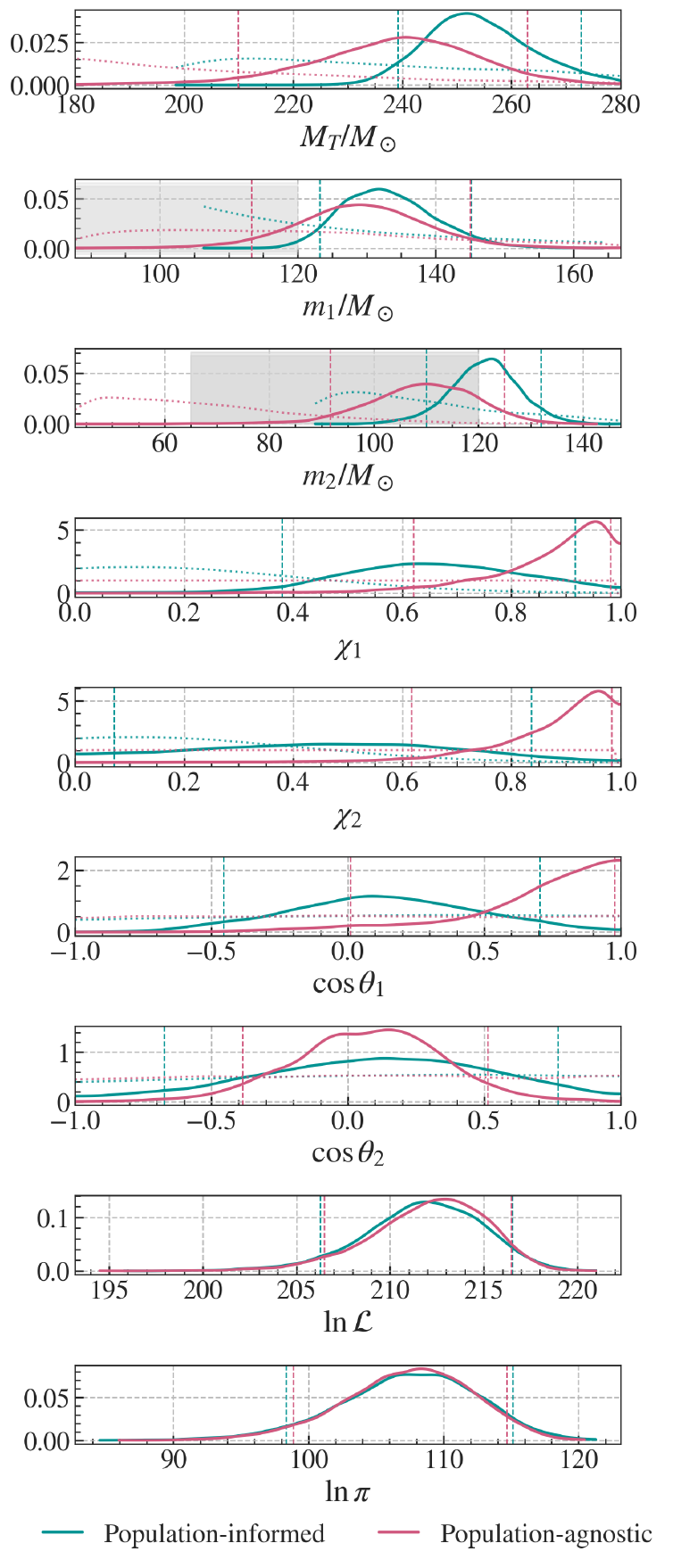}
    \caption{Posterior distributions (solid lines) and prior distributions (dotted lines) for binary parameters of GW231123, obtained using population-agnostic (red) and population-informed (teal) non-eccentric analyses. The vertical dashed lines in each panel mark the 5th and 95th percentiles of the posterior, enclosing the 90\% credible interval. For each analysis, the prior is shown over the same parameter range as the posterior. Where the prior has no support due to hard boundaries imposed by the prior, the dotted curve terminates abruptly, reflecting a true absence of prior probability rather than a plotting artefact. The grey-shaded regions in the component mass panels indicate the range commonly associated with the pair-instability supernova gap, within which stellar evolution models predict a dearth of black hole formation.}
    \label{fig:population-informed}
\end{figure}

As a final check, we assess \textit{how unusual GW231123 is relative to the observed \ac{bbh} population} by reanalysing the data using the  \nrsur{} waveform model using two different non-eccentric priors: 
\begin{inparaenum}[(1)]
    \item a population-agnostic prior with choices similar to \ac{lvk}.
    \item an empirically obtained population-informed prior that reflects current astrophysical knowledge of \ac{bbh} demographics. 
\end{inparaenum}
The population-informed prior is consistent with the \textsc{Default} \ac{bbh} population model adopted in GWTC-4.0, with all hyperparameters fixed to their median values inferred from the catalogue~\citep{LIGOScientific:2025pvj}. We note that GW231123 is included in this catalogue; however, in hierarchical inference, each event contributes a marginal likelihood term weighted by the population model evaluated at its parameters. Since GW231123 sits in the tail of the mass distribution, this term is intrinsically small and varies slowly with the hyperparameters. Combined with being one of \(> 150\) events in the catalogue, its influence on the inferred hyperparameters is negligible. In this model, the primary source-frame mass \(m_1\) is drawn from a mixture distribution comprising a doubly truncated power law and two Gaussian components, with peaks near \(\sim 10\,M_\odot\) and \(\sim 35\,M_\odot\), and a smooth low-mass turn-on at \(\sim5\,M_\odot\). The mass-ratio \(q=m_2/m_1\) distribution follows a power law, and shares the same minimum mass and smoothing scale.

The component spin magnitudes \(\chi_{1,2}\) follow a truncated Gaussian distribution, while the spin tilt angles are described by a mixture of a truncated Gaussian and an isotropic component. The catalogue data favour modest spin magnitudes preferentially aligned with the orbital angular momentum, consistent with expectations from isolated binary evolution scenarios \citep{Mandel:2018hfr, Mapelli:2020vfa}; it is this data-informed character of the hyperparameters, rather than any assumption built into the prior structure itself, that suppresses the large in-plane spins originally inferred for this event under a population-agnostic prior.

Figure~\ref{fig:population-informed} shows the marginal posteriors for different binary parameters such as the total source-frame mass \(M_T\), component source masses \(m_{1,2}\), component spin magnitudes \(\chi_{1,2}\), and spin-tilt cosines \(\cos\theta_{1,2}\), inferred using our priors. Independent of the prior choice, both analyses robustly infer an exceptionally massive system, with \(M_T \gtrsim 200\,M_\odot\) at 90\% credibility, placing the source well above the bulk of the empirically inferred population~\citep{LIGOScientific:2025pvj, Mould:2026nle}. Notably, the population-informed prior favours an even larger total mass and is in tension with the conclusions of \citet{Mandel:2025qnh}, who rely on prior reweighting. Additionally, we find that regardless of the prior choices, the heavier component is highly likely to straddle the pair-instability gap (shown in grey), whereas the secondary component is most likely to lie within the gap.

In contrast to the mass parameters, the inferred spin properties depend strongly on the assumed prior. Under the population-agnostic prior, the posteriors support large spin magnitudes--particularly for the primary, with \(\chi_1 \gtrsim 0.7\)--and allow a broad range of spin-tilt angles, including substantial misalignment.
The population-informed prior instead yields lower spin magnitudes for both components and strongly suppresses support for large in-plane spins, resulting in tighter posteriors concentrated toward moderate spins and more aligned configurations. This is consistent with the findings of \citet{Mould:2026nle}, who report a reduction in the primary spin magnitude when the catalogue is jointly analysed and with \citet{Wolfe:2026dcq}, who report a noticeable impact on the inferred population spin distribution when including GW231123.
However, this suppression reflects the data-informed hyperparameters of the parameterised population model rather than any intrinsic feature of the prior structure.
Non-parametric population models yield greater support for near-extremal spins; nonetheless, as their inferred spin magnitude distributions also fall steeply toward \(\chi \to 1\), the population-agnostic prior --- being uniform in spin magnitude --- retains better support at near-extremal spins than any current population-informed prior, parametric or otherwise.

Posterior samples alone, however, do not establish the probabilistic consistency of an analysis. We therefore compare the model evidence and find that the population-agnostic prior is favoured over a population-informed one with \(\log_{10} \mathcal{B}\sim 1.29\). We emphasise that the prior ranges used are identical between the two analyses --- only the prior shape differs --- so this preference cannot be attributed to differences in prior volume. We do not interpret this as evidence that the population model is incorrect; rather, it quantifies the tension between the data and a prior reflecting the bulk of the current \ac{bbh} population, and some preference for the agnostic prior is expected for a tail event. To understand the origin of this preference, we examine the posterior distributions of the log-prior and log-likelihood as diagnostics; these are distinct from the evidence itself, which is the likelihood averaged over the prior. The near-identical \(\ln\pi\) distributions confirm that prior volume is not driving the Bayes factor. The \(\ln\mathcal{L}\) distribution instead shifts toward higher values under the population-agnostic prior, indicating that the population-informed prior restricts access to the high-likelihood region of parameter space. This highlights the risk of drawing conclusions based solely on posterior samples: a posterior that appears more ``astrophysically plausible'' does not necessarily correspond to a better explanation of the data. It also illustrates a key limitation of maximum-likelihood reasoning, which, in this case, would incorrectly suggest equal preference between the two analyses.

\section{Conclusion}
\label{sec:conclusion}

Our analysis shows that at \acp{snr} \(\gtrsim 20\), it is possible to distinguish between quasi-spherical and eccentric signal hypotheses for GW231123 using indistinguishability criteria and model-selection diagnostics, when appropriate waveform models are employed.
At the same time, we find that strong degeneracies can arise in parameter estimation when incomplete waveform models are used. Numerically simulated eccentric \ac{bbh} signals with modest component spins can be recovered as quasi-spherical systems with large in-plane spin components, as missing eccentric features are partially absorbed. 
This is consistent with previous results reported in \citet{CalderonBustillo:2020xms, Divyajyoti:2025cwq, Romero-Shaw:2022fbf}, but shows for the first time that near-extremal spins can be induced by lack of eccentric corrections in the waveform template.
Instead, we find that ringdown-only analyses of simulated signals can robustly recover the remnant parameters, if an unbiased estimate of the signal peak is known.
Recovery relies on the longest-lived \ac{qnm}, since at current sensitivity we cannot extract evidence for the presence of multiple modes within the \ac{qnm} validity regime.

We quantify the astrophysical consequences of such modelling assumptions through full Bayesian analyses and complementary post-peak studies of GW231123. 
Our \ac{imr} analysis using eccentric \ac{nr} templates yields statistically acceptable fits to the data, with residuals consistent with detector noise, while favouring solutions with larger remnant masses compared to \nrsur{}, and large initial eccentricities. However, quasi-spherical \nrsur{} models achieve higher maximum-likelihood values, owing to their ability to reproduce the signal's dominant peak morphology.
Post-peak studies infer properties consistent with \nrsur{} when adopting the most conservative estimate of the signal peak.
However, at earlier times they instead point to remnant properties consistent with eccentric templates and in tension with \nrsur{}, possibly pointing to a secondary likelihood peak that cannot be accessed under the quasi-spherical hypothesis.

We also find that, consistent with past analyses~\citep{Mandel:2025qnh, Tenorio:2026dcc}, inferred spins under the quasi-spherical hypothesis are reduced when using population-informed parametric priors rather than population-agnostic ones. However, model selection disfavors the population prior with a Bayes factor \(\mathcal{B} > 19\), underscoring the exceptional nature of this event within the current \ac{bbh} population.

Collectively, these results indicate that while eccentric configurations can provide viable descriptions of the data, they are not favoured. In particular, highly eccentric (\(e_0 \gtrsim 0.5\)) merger scenarios are disfavoured relative to quasi-spherical interpretations, which provide a better overall description of GW231123.\\

\section{Acknowledgements}
The authors thank Rossella Gamba and Matthew Mould for helpful comments and suggestions. KC particularly thanks Rossella for motivating him to complete this work and also thanks Ítalo Ferreira for encouragement during the work. 
The authors acknowledge the use of an OpenAI language model (ChatGPT) for assistance in improving the presentation of figures.
This research has made use of data obtained from the Gravitational Wave Open Science Center (\href{https://gwosc.org}{gwosc.org}), a service of the LIGO Scientific Collaboration, the Virgo Collaboration, and KAGRA. This material is based upon work supported by NSF's LIGO Laboratory, a major facility fully funded by the National Science Foundation, together with support from the Science and Technology Facilities Council (STFC) of the United Kingdom, the Max Planck Society (MPS), and the State of Niedersachsen/Germany for the construction of Advanced LIGO and the construction and operation of the GEO600 detector. Additional support for Advanced LIGO was provided by the Australian Research Council. The authors acknowledge the use of the Gwave cluster at Pennsylvania State University for the computational analyses presented here (supported by NSF grants OAC-2346596, OAC-2201445, OAC-2103662, OAC-2018299, and PHY-2110594).

\section{Data Availability Statement}

The \ac{nr} waveforms analysed in this work are publicly available through the \href{https://data.black-holes.org/simulations/index.html}{SXS}, \href{https://ccrg.rit.edu/numerical-simulations}{RIT}, and \href{https://egrav.icc.ub.edu/}{ICCUB} waveform catalogues. The gravitational-wave strain data for GW231123 used in this study are available through the Gravitational Wave Open Science Centre \citep{LIGOScientific:2026jgl} and can be accessed \href{https://gwosc.org/eventapi/html/O4_Discovery_Papers/GW231123_135430/v1/}{here}. Most of the analysis scripts and code used to generate the figures in this work are available on \href{https://github.com/koustavchandra/GW231123_eccentric_analysis}{GitHub}. Others will be made available on reasonable request. No new observational data were generated as part of this study.

\clearpage

\bibliographystyle{mnras}
\bibliography{bibliography}

\end{document}